\let\pdfoutput\defined
\documentclass[cits]{PoS}

\usepackage{amsmath,amsfonts,amssymb,mathrsfs}

\title{%
\vspace*{-1cm}
\begin{minipage}{\textwidth}
\begin{flushright}
\texttt{\footnotesize 
PoS(LAT2006)037\\%
hep-lat/0609042\\%
BNL-HET-06-12\\%
DESY-06-172\\% 
} 
\end{flushright}
\end{minipage}\\[15pt]   
Multi-Step stochastic correction in dynamical fermion updating algorithms}

\ShortTitle{Multi-Step stochastic correction in dynamical fermion updating algorithms}
\author{\speaker{Enno E. Scholz}\\
  Brookhaven National Laboratory, Upton, NY 11973, USA\\
  E-mail: \email{scholzee@quark.phy.bnl.gov}}
\author{Istv\'an Montvay\\
  Deutsches Elektronen-Synchrotron---Theory Group, Notkestr. 85, 22603 Hamburg, Germany\\
  E-mail: \email{istvan.montvay@desy.de}}

\abstract{%
The advantages of using Multi-Step corrections for simulations of lattice
gauge theories with dynamical fermions will be discussed. This technique is
suited for algorithms based on the Multi-Boson representation of the dynamical
fermions as well as for the Hybrid Monte-Carlo (HMC) algorithm and variants of
the latter, like the Polynomial-HMC. Especially the latter has the power to
deal with an odd number of fermion fields---an essential feature necessary for
realistic QCD-simulations with up-, down-, and strange-quarks. In particular,
we will discuss the application of the multi-step (actually two-step) correction technique to a PHMC updating algorithm for twisted-mass Wilson fermions with non-degenerate fermion masses, as it was used in recent dynamical simulations for $N_f=2+1+1$ fermion flavors. 
}

\FullConference{XXIVth International Symposium on Lattice Field Theory\\
                July 23-28, 2006\\
                Tucson, Arizona, USA}
\def\nicefrac#1#2{\leavevmode\kern.1em\raise.5ex\hbox{\the\scriptfont0 #1}\kern-.1em/\kern-.15em\lower.25ex\hbox{\the\scriptfont0 #2}}

\begin{document}

%%%%%%%%%%%%%%%%%%%%%%%%%%%%%%%%%%%%%%%%%%%%%%%%%%%%%%%%%%%%%%%%%%%%%%%%
%%%%%%%%%%%%%%%%%%%%%%%%%%%%%%%%%%%%%%%%%%%%%%%%%%%%%%%%%%%%%%%%%%%%%%%%
%
%
%
%moved to abstract, but PoS didn't like that...
%
\noindent This article is organized as follows: in the first part we will briefly review the idea of using Multi-Step stochastic correction, for more details we refer to our paper \cite{Montvay:2005tj}. We provide a description of the \emph{Polynomial Hybrid Monte-Carlo algorithm with stochastic correction} , which was successfully applied in \cite{Chiarappa:2006ae} to simulate $N_f=2+1+1$ quark flavors using the twisted-mass fermion action. In the remainder we will summarize some of the results of these simulations, again for more details we refer to \cite{Chiarappa:2006ae}.

%%%%%%%%%%%%%%%%%%%%%%%%%%%%%%%%%%%%%%%%%%%%%%%%%%%%%%%%%%%%%%%%%%%%%%%%
%%%%%%%%%%%%%%%%%%%%%%%%%%%%%%%%%%%%%%%%%%%%%%%%%%%%%%%%%%%%%%%%%%%%%%%%

\section{\label{sec:stochCorr}Multi-Step stochastic correction}

The idea of Multi-Step stochastic correction is a natural extension of the
Two-Step technique well established for Multi-Boson algorithms (TSMB-alg.),
cf.\ \cite{Montvay:1995ea,Farchioni:2002vn} and references
therein. One uses multiple correction steps with increasing precision. Our
choice to control the precision in each step is to use polynomials $P_i$ of increasing order $n_i$ (see Sec.\ \ref{sec:polyApprox}) to approximate the required power in the pseudo-fermionic part of the action:
%
%\vspace{-.1cm}
\begin{equation}
P_1(\tilde Q^2)\;\simeq\;(\tilde Q^2)^{-\alpha},\;\;\;P_i(\tilde Q^2)\;\simeq\;\Big[(\tilde Q^2)^\alpha P_1(\tilde Q^2)\cdots P_{i-1}(\tilde Q^2)\Big]^{-1},
\end{equation}
%\vspace{-.1cm}
% 
where $\tilde Q$ denotes the Hermitian fermion-matrix (containing either one
fermion-flavor or in case of the twisted mass formulation one flavor
doublet). For example, $\alpha=\nicefrac12$ results in a one flavor (twisted-mass: one
flavor-doublet) action, $\alpha=1$ leads to $N_f=2$ (twisted-mass: two degenerate flavor-doublets).   

The update-sequence now consists of several nested noisy corrections
(cf. \cite{Farchioni:2002vn} and references therein) using the various
polynomials.\footnotemark As an example we show a sequence using three-step correction: 
\footnotetext{For the noisy correction step we also need polynomials $\bar P_i(x)\simeq P_i(x)^{-1/2}$ of order $\bar n_i$ to generate the noise vector from a Gaussian distribution.}
\begin{equation}
\mbox{start}\:\underbrace{\underbrace{\underbrace{\rightarrow\cdots\rightarrow}_{N_1\mbox{\footnotesize
   steps} P_1}{P_2}\longrightarrow\cdots\longrightarrow
   \underbrace{\rightarrow\cdots\rightarrow}_{N_1\mbox{\footnotesize
   steps} P_1}{P_2}}_{{N_2\mbox{\footnotesize
   steps}}}\:\longrightarrow\:{P_3}\longrightarrow\underbrace{\cdots{\cdot}\cdots{\cdot}}
_{(N_1\,N_2)\times
   P_1,
   {N_2\times P_2}}\longrightarrow\:{P_3}}_{{N_3\mbox{\footnotesize
   steps}}}
\end{equation}
Each time a configuration is rejected it will be replaced with the last accepted one by that same correction step.

In the following subsections, we describe applications of this technique to Multi-Boson as well as Hybrid Monte-Carlo based algorithms.

%%%%%%%%%%%%%%%%%%%%%%%%%%%%%%%%%%%%%%%%%%%%%%%%%%%%%%%%%%%%%%%%%%%%%%%%%

\subsection{Multi-Step Multi-Boson algorithm}

The update for the first (Multi-Boson) step using $P_1$ as well as for the gauge-fields are exactly the same as for the TSMB-algorithm as described in \cite{Farchioni:2002vn} and applied in \cite{Farchioni:2003bx, Farchioni:2003nf, Farchioni:2004tv, Farchioni:2004us}, we just add more correction-steps to the procedure. This changes the numerical cost expressed in units of \emph{matrix-vector-multiplications (MVMs)} given for TSMB in \cite{Farchioni:2002vn} (for unexplained notation confer that publication) to
\begin{equation}
\frac{\textrm{cost}}{\textrm{MVM}}\;=\;6(n_1N_\phi+N_U)+I_GF_G+\sum_{k=2}^i(n_k+\bar n_k)N_{Ck},
\end{equation}
$N_{Ck}$ is the number of times the correction step involving $P_k$ is called. In \cite{Montvay:2005tj} we gave an example how calling the most expensive correction step less often actually works in a numerical simulation.

%%%%%%%%%%%%%%%%%%%%%%%%%%%%%%%%%%%%%%%%%%%%%%%%%%%%%%%%%%%%%%%%%%%%%%%%%%%

\subsection{\label{sec:hmcphmc}HMC and PHMC with stochastic correction}

The first update-step does not necessarily have to rely on a Multi-Boson
update. It can for instance be replaced by a standard \emph{Hybrid Monte-Carlo(HMC)}-update. The advantage of combining stochastic corrections with HMC algorithms is that one may use a mass-preconditioned fermion-matrix $\tilde Q^2+\mu_1$ in the basic update step, and account for this in the stochastic correction steps, using there subsequently decreasing $\mu_i$  to achieve the ``target action'' ($\mu_N=0$) in the last ($N^{\textrm{th}}$) step ($\mu_i\in\mathbb{R}$, $\mu_1>\ldots>\mu_i>\ldots>\mu_N=0$).

A severe limitation of the traditional HMC algorithm is, that only an even
number of fermionic fields can be simulated. To overcome this drawback one has
to take a fractional power of the squared fermion matrix, which can be done by
using an approximation of the function $x^{-\alpha}$, e.g.\
$\alpha=\nicefrac12$. Commonly, either a \emph{polynomial} or a \emph{rational}
approximation is used, leading to a so-called PHMC or RHMC algorithm,
respectively. The PHMC was suggested in
\cite{deForcrand:1996ck,Frezzotti:1997ym,Frezzotti:1998eu,Frezzotti:1998yp},
while a description of the RHMC can be found in \cite{Clark:2006fx}. The
original proposal of the PHMC algorithm had to face the problem, that large
orders in the polynomial approximation forbid an efficient application to the
region of light quark masses. In the remainder of this section we will
describe our approach to introduce a stochastic correction step to make this
algorithm competitive to other algorithms currently used. For results on a
successful application see Sec.\ \ref{sec:tmtwo11}.  An alternative to the
stochastic correction would be to reweight the generated configurations, this is currently investigated by Chiarappa \textit{et al.}, see \cite{Chiarappa:2005mx}.

The HMC algorithm and its variants all start by introducing a fictitious time-scale $\tau$ and evolve the gauge and conjugate momenta fields according to Hamilton's equations of motion. Here we will only describe the details concerning the polynomial approximation and correction step in the fermionic part of the action. Again, we use a polynomial approximation $P_1({\tilde Q}^2)\:\simeq\: (\tilde Q^2)^{-\alpha}$ of order $n_1$,
%%
%\begin{equation}
%P_1({\tilde Q}^2)\:\simeq\: (\tilde Q^2)^{-\alpha},
%\end{equation}    
%%
where $P_1$ may be evaluated in a recursive scheme (which will be convenient
in the Metropolis step) or by using the root-representation
\begin{equation}
P_1({\tilde Q}^2)\:=\:c_0 \prod_{i=1}^{n_1} ( {\tilde Q}^2 - r_i )\:=\:c_0 \prod_{i=1}^{n_1}( {\tilde Q} -\rho_i ) \, \prod_{i=n_1}^1 ( {\tilde Q} - \rho_i^* ).
\end{equation}
The last rearrangement is possible because for $n_1$ even (to which we restrict ourselves) the roots always appear in complex conjugate pairs. Note the ordering in the two products (first increasing index, last decreasing index): this assures a better numerical stability (the roots $r_i$ itself are also ordered to a scheme to minimize the numerical error \cite{Montvay:1997vh})
 and allows for a efficient way of calculating the fermionic force, as needed in the HMC-step. By introducing the auxiliary fields
\vspace{-.10cm}
\begin{eqnarray}
\phi_1^{(k)}&\equiv&\sqrt{c_0}\,\phi\,(\tilde Q - \rho_1)\cdots(\tilde Q - \rho_k),\;\;\;k=1,\ldots,n-1\\
%                  &=& \phi_1^{(k-1)}\,(\tilde Q - rho_k)\\
\phi_2^{(k)}&\equiv&\sqrt{c_0}\,\phi(\tilde Q -\rho_1)\cdots(\tilde Q - \rho_n)(\tilde Q-\rho_n^*)\cdots(\tilde Q-\rho_{k+2}^*)
\vspace{-.10cm}
\end{eqnarray}
and setting $\phi_1^{(0)}=\sqrt{c_0}\phi$, the fermionic force can be written as
\begin{eqnarray}\nonumber
\phi D_{x\mu j} P_1(\tilde Q^2) \phi^\dagger &=& \sum_{k=0}^{n-1}\Big[ {\phi_1^{(k)}}\,(D_{x\mu j}\tilde Q)\,{\phi_2^{(k)\dagger}}\:+\:{\phi_2^{(k)}}\,(D_{x\mu j}\tilde Q) {\phi_1^{(k)\dagger}} \Big] %\\
% &=& 
\:=\:2 \textrm{Re}\left(\sum_{k=0}^{n-1} {\phi_1^{(k)}}\,(D_{x\mu j} \tilde Q) {\phi_2^{(k)\dagger}}\right).\\
\label{eq:fermForce}
\end{eqnarray}
Here $\phi$ always denotes the pseudo-fermionic fields from the
``bosonization'' of the fermion matrix and $D$ is the derivative with respect
to the given indices. Observe, that it is most efficient to first calculate all the $\phi_1^{(k)}$ and store the results. In the actual computation one than has to combine the actual $\phi_2^{(k)}$ (which is obtained recursively from the previous one) with one of the prepared $\phi_1^{(k)}$-fields. Using a better conditioned fermion matrix by applying \emph{even-odd-preconditioning} is also possible, in that case a similar scheme can be written down.

After a PHMC-step, which starts with generating the pseudo-fermionic fields according to $\phi\:=\:\bar P_1(\tilde Q^2)\,\eta$ from a Gaussian-distributed vector $\eta$ using a polynomial $\bar{P}_1(x)\simeq P_1(x)^{-1/2}$ of order $\bar n_1$ %
%%
%\[\phi\:=\:\bar P_1(\tilde Q^2)\,\eta\]
%%
and the real conjugate momenta field $P_{x\mu j}$ (in the adjoint
representation) according to the contribution $\textrm{d}P\exp(-P^2/2)$, a
trajectory of length $\delta\tau$ is performed using the Sexton-Weingarten
integration scheme \cite{Sexton:1992nu} with multiple time-scales for the
gauge and fermionic parts. At the end of every trajectory a global
accept-reject step is performed, using the polynomial $P_1$ in its recursive
form to calculate the fermionic part of the energy. The reason for this is to
correct for numerical errors originating from the finite step size in the
integrator. After $N_{\textrm{traj}}$ such trajectories a stochastic
correction step is carried out as described in Sec.\ \ref{sec:stochCorr},
using polynomials $P_2$ and $\bar P_2$ in the recursive representation. In
this way we are able to correct for the (intentionally) poor approximation in
the PHMC-step, which allows us to prepare trajectories at moderate numerical cost. An approximate formula\footnotemark\ for the cost in number of matrix-vector-multiplications (\emph{MVMs}) can be given as follows:
\footnotetext{Here we neglected the cost from calculating the force from the gauge and conjugate moment parts and made the assumption that the tensor-like multiplication in Eq.\ (\ref{eq:fermForce}) counts as one MVM.} 
\begin{equation}
\frac{\textrm{cost}}{\textrm{MVM}}\approx 2n_B(n_2+\bar n_2)\:+\:N_{\textrm{traj}}\Big[2n_B(n_1+\bar n_1)\,+\,n_B(3+2N_Q)(4n_1-1)\Big],
\end{equation}
where $n_B$ is the number of determinant-breakup used (see below) and $N_Q$ denotes the number of time-steps for the fermionic integration.

%%%%%%%%%%%%%%%%%%%%%%%%%%%%%%%%%%%%%%%%%%%%%%%%%%%%%%%%%%%%%%%%%%%%%%%%%

\subsection{Determinant-breakup, mass preconditioning}

An important improvement of the Multi-Boson algorithms and the PHMC- as well as the RHMC-algorithms is so-called ``determinant-breakup'' \cite{Hasenbusch:1998yb}. The fractional powers in the approximation allow to use $n_B$ sets of pseudo-fermion fields with $\alpha$ replaced by $\alpha/n_B$. This technique was proven to improve the simulations for the TSMB- \cite{Farchioni:2003bx, Farchioni:2003nf, Farchioni:2004tv, Farchioni:2004us} and the PHMC- \cite{Chiarappa:2006ae} and also recently for the RHMC-algorithm \cite{Clark:2006fx}.

Other techniques to speed up the simulation rely on preconditioning of the fermion-matrix, we already mentioned mass-preconditioning briefly in Sec.\ \ref{sec:hmcphmc}. Details on how to use this in conjunction with multiple correction steps are given in \cite{Montvay:2005tj}.

%%%%%%%%%%%%%%%%%%%%%%%%%%%%%%%%%%%%%%%%%%%%%%%%%%%%%%%%%%%%%%%%%%%%%%%%%

\subsection{\label{sec:polyApprox}Polynomial approximation}

There are several different methods available for the required approximations
in the MSMB and HMC-variant algorithms. Most commonly used are
\emph{polynomial} or \emph{rational} approximations of some order $n$. Usually
for a given precision the rational approximation requires a lower order
\cite{Clark:2006fx}, but here we want to stress the fact that while the order
of a polynomial determines the cost in matrix-vector-multiplications directly,
for the rational function one has to invert the fermion matrix at least once,
which is the main contribution to the cost and is itself a polynomial
approximation, too. Therefore we prefer to take a direct polynomial
ansatz.\footnotemark

\footnotetext{Note, that although the polynomial degrees of the first
  approximation steps are fixed, the degree of the last (and most expensive)
  step can be adaptive because of the recursive evaluation.}

\begin{figure}[t]
\begin{center}
\includegraphics[angle=-90, width=.32\textwidth]{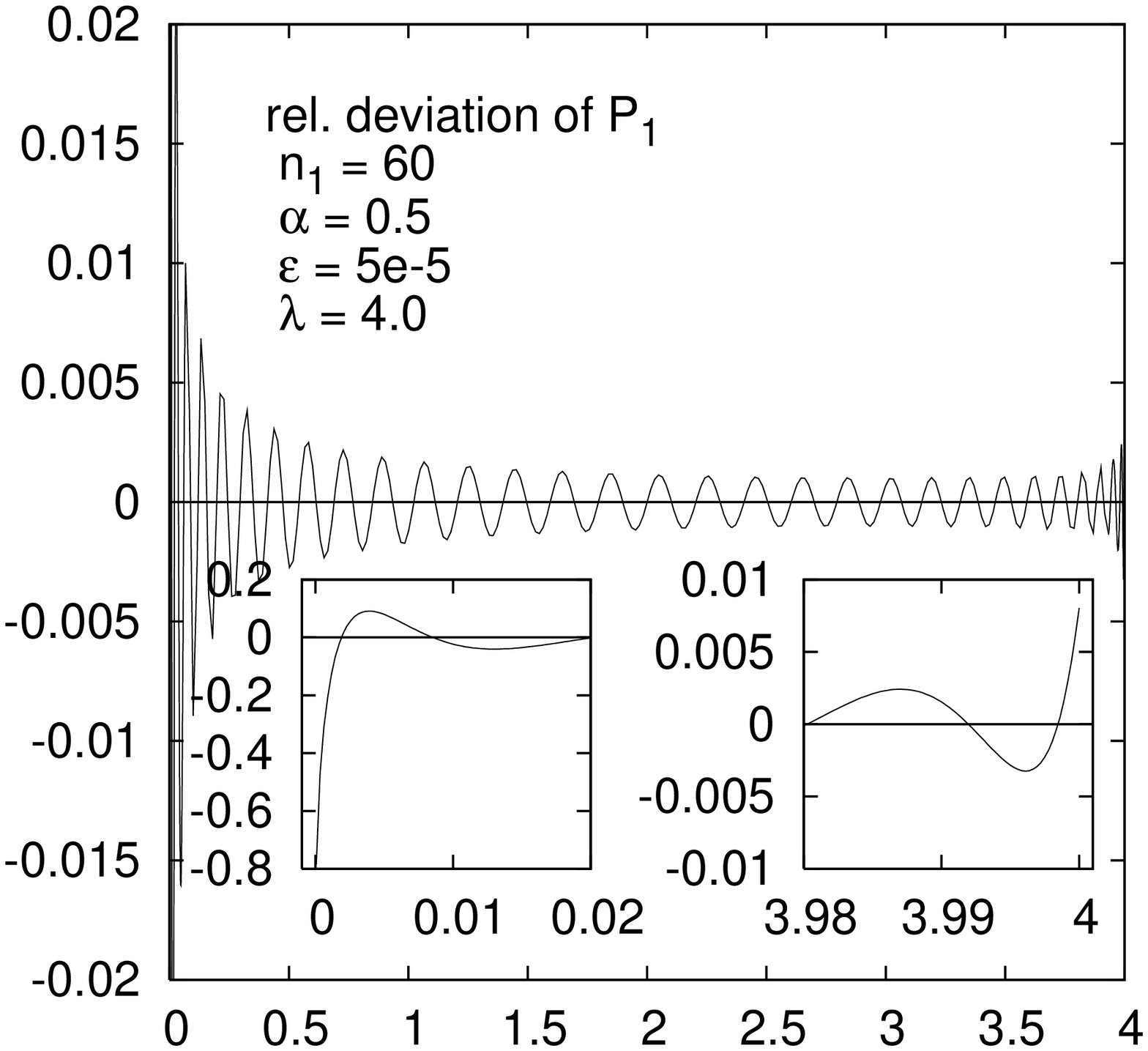}%
\includegraphics[angle=-90, width=.32\textwidth]{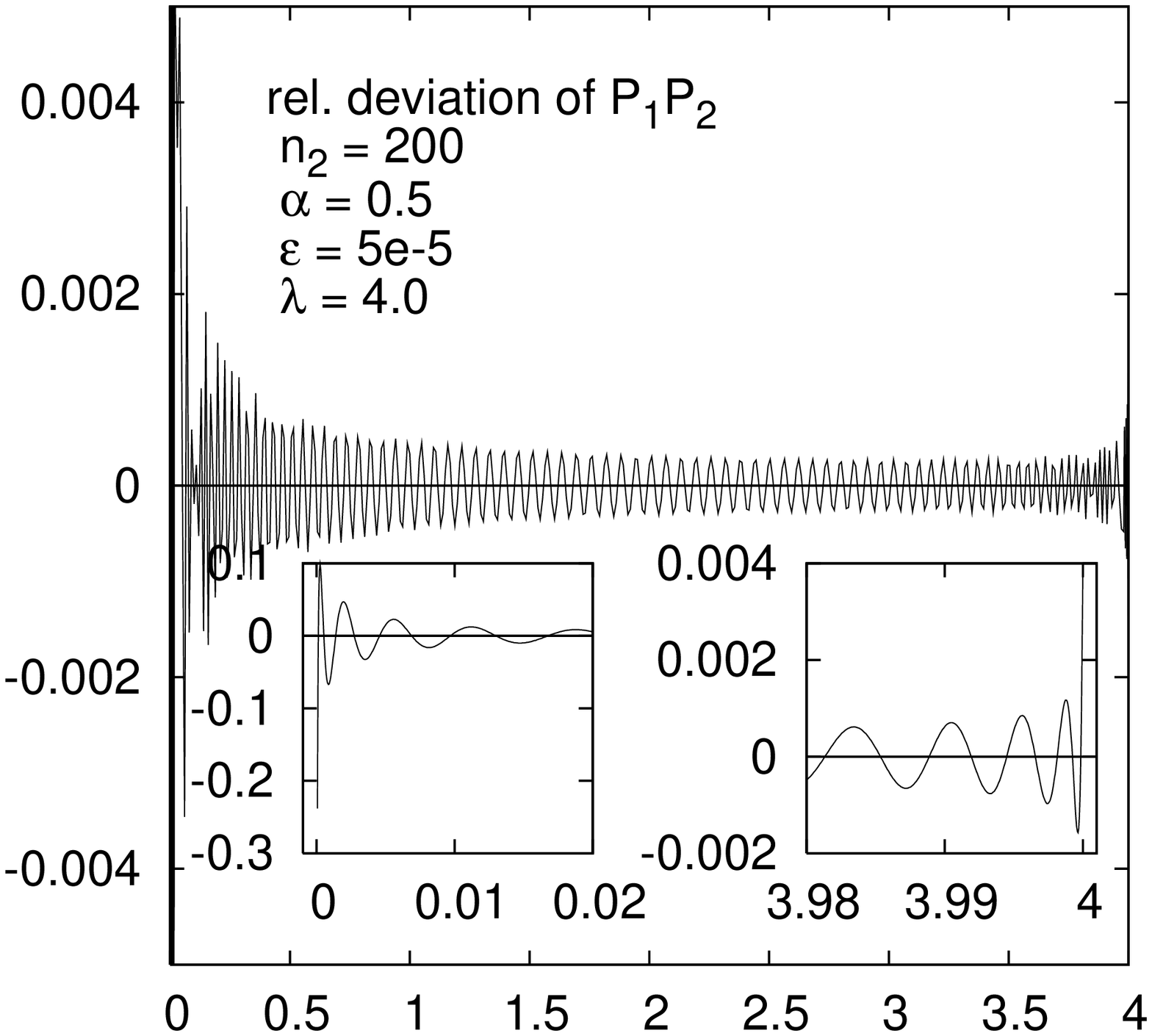}%
\includegraphics[angle=-90, width=.32\textwidth]{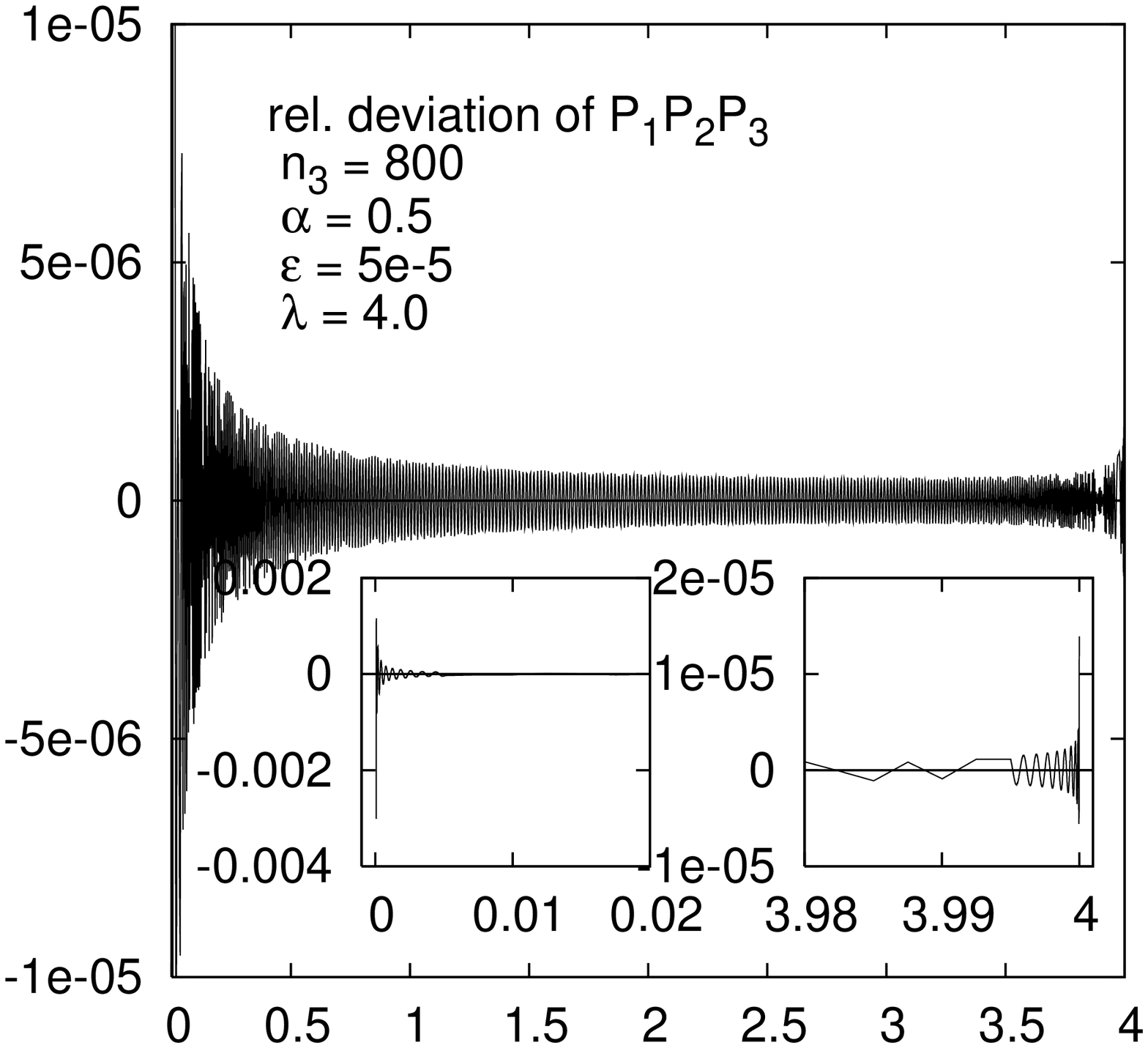}%
\end{center}
\vspace*{-1.00cm}
%\vspace*{-1.2cm}
\caption{Example for polynomials with increasing order to approximate $x^{-1/2}$ in the interval $[5\cdot10^{-5}, 4.0]$ as used in a MSMB-update \cite{Montvay:2005tj}, shown is the relative deviation}
\label{fig:polynoms}
%\vspace*{-.65cm}
\end{figure}

Another question arises, namely in which way the error of the approximation is
minimized. In principle, defining any norm and calculate the deviation
according to it is a legitimate procedure. Commonly used are either the $L_2$-
or the $L_\infty$-norm. The first minimizes the quadratic deviation, while the
latter minimizes the maximal deviation. It has been shown that the first one
is better suited for dynamical fermion simulations \cite{Montvay:1999kq},
since it leads to smaller deviations in the bulk of the
approximation-interval at the cost of a higher deviation at the
boundaries. The $L_\infty$-norm leads to an uniformly distributed deviation,
which is disadvantageous because most of the eigenvalues of the fermion matrix
are in the bulk and only a few ones are close to the lower boundary. Examples
of the $L_2$-optimized polynomial approximations as used in a three-step multi-boson correction update \cite{Montvay:2005tj} are shown in Fig.\ \ref{fig:polynoms}. For the generation of the coefficients we refer to \cite{Montvay:1997vh,Gebert:2003pk,Katz:2004ki}.   

%%%%%%%%%%%%%%%%%%%%%%%%%%%%%%%%%%%%%%%%%%%%%%%%%%%%%%%%%%%%%%%%%%%%%%%%%%%
%%%%%%%%%%%%%%%%%%%%%%%%%%%%%%%%%%%%%%%%%%%%%%%%%%%%%%%%%%%%%%%%%%%%%%%%%%%

\section{\label{sec:tmtwo11} Twisted mass-simulations with $\boldsymbol{N_f=2+1+1}$}

As part of the efforts of the \emph{European Twisted Mass-Collaboration (ETM-Coll.)}, the PHMC-algorithm with one stochastic correction step as
introduced above has been applied to simulate two doublets of twisted mass
quarks \cite{Chiarappa:2006ae}. In the twisted mass action \cite{Frezzotti:2000nk} an additional mass
term is added, which results in a lower bound for the eigenvalue spectrum and,
if the standard mass term is properly tuned, in an automatically
$\mathcal{O}(a)$-improved fermion action. The first doublet contains two
mass-degenerate light quark flavors (up and down), whereas in the second
doublet a mass-splitting term as introduced by Frezzotti and Rossi
\cite{Frezzotti:2003xj} has been added. Therefore, the quarks of the second
doublet can be identified as the charm and strange flavors, having different
masses. In the mass-split case we have to use an algorithm which is capable
of simulating a single fermion flavor. In principle it would have been
possible to use the HMC as in \cite{Urbach:2005ji} for the first doublet and
  the PHMC only for the second one. For simplicity we have chosen to use the
  PHMC-algorithm for both doublets and the results showed that the PHMC with
  stochastic correction works fine for light quarks, too.  

We performed simulations at two lattice spacings on a fixed physical volume of
$aL\simeq2.4\textrm{fm}$ ($a\approx0.20\textrm{fm}$ at $L^3\times
T=12^3\times24$ and $a\approx0.15\textrm{fm}$ at $16^3\times32$) using the
Symanzik tree-level improved gauge action \cite{Symanzik:1983dc}. In both
cases the twisted mass parameters where kept fixed in physical units and the
untwisted mass parameter was varied. On the coarse lattice the numerical cost\footnotemark\ to generate a new configuration varied between $28.2$ and $46.5$ (in thousands of MVMs) for the heaviest and lightest pion mass. On the finer lattice we observed similar numbers between $17.9$ and $44.2$. For more details see \cite{Chiarappa:2006ae}. That allowed us to study the phase structure and determine the critical value for the untwisted mass parameter, which is important to obtain the $\mathcal{O}(a)$-improvement.

%\footnotetext{These configurations are not independent. For a detailed cost-analysis the autocorrelation, which depends on the measured quantity, has to be taken into account.}

\footnotetext{To compare these numbers with those from other groups: $1 \textrm{MVM} \simeq 2688\cdot \Omega\,\textrm{flop}$, where $\Omega=L^3\times T$ is the lattice-volume.}

The volume scaling of the algorithm is very good: keeping the action parameters constant, for instance, between $12^3 \times 24$ and $24^3 \times 48$ lattices, one can keep the polynomial degrees $n_2$, $\bar{n}_2$ constant and increase $(n_1+\bar{n}_1)$ only by about 10{\%} for the same lower bound of the approximation intervall. Even this moderate increase can be compensated by increasing the lower bound which is possible due to the diminished fluctuations of the smallest eigenvalues.

%%%%%%%%%%%%%%%%%%%%%%%%%%%%%%%%%%%%%%%%%%%%%%%%%%%%%%%%%%%%%%%%%%%%%%%%%%%

\subsection{\label{sec:physRes}Results}

\begin{figure}[t]
\begin{center}
\includegraphics[angle=-90, width=.32\textwidth]{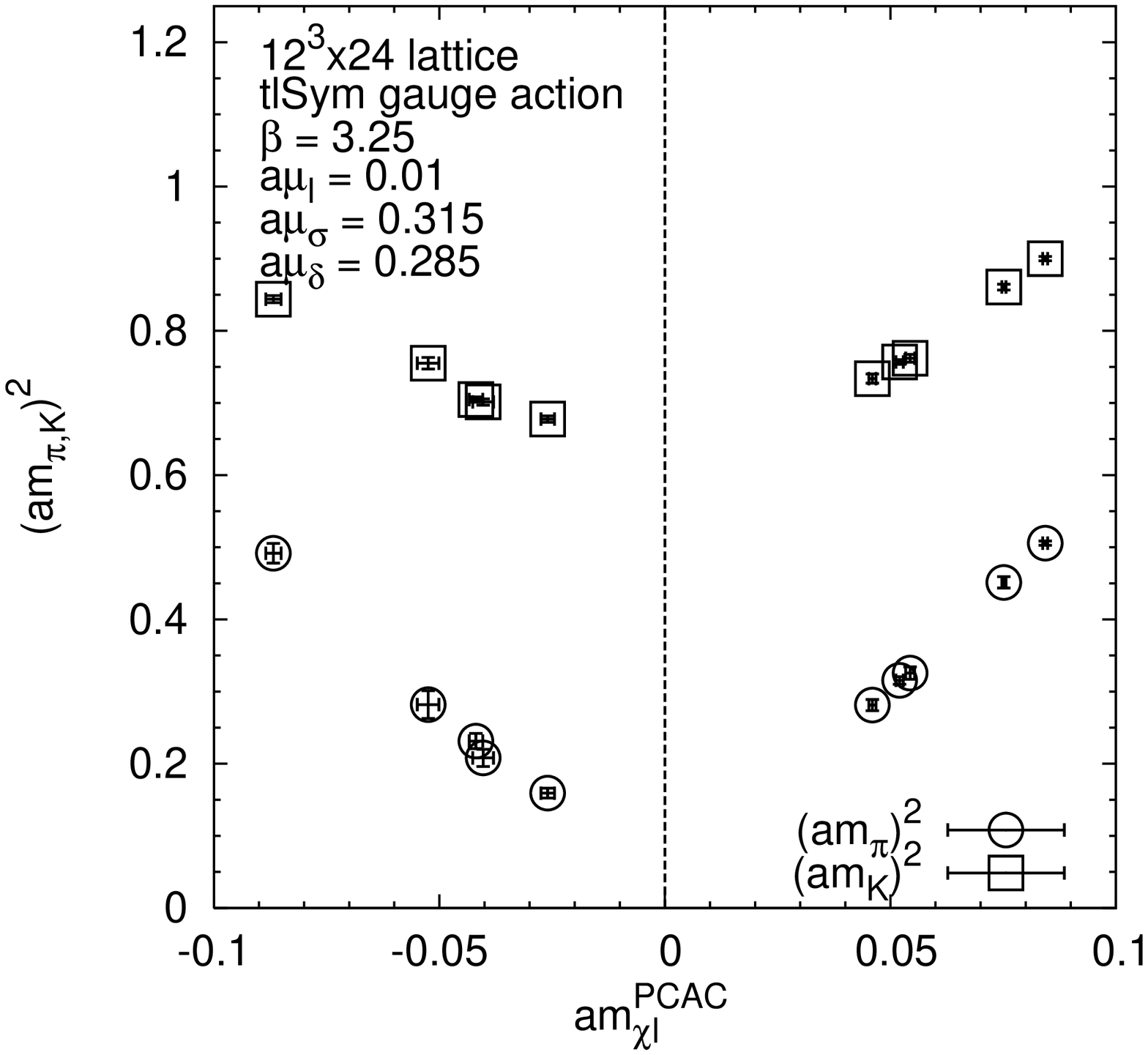}
\includegraphics[angle=-90, width=.32\textwidth]{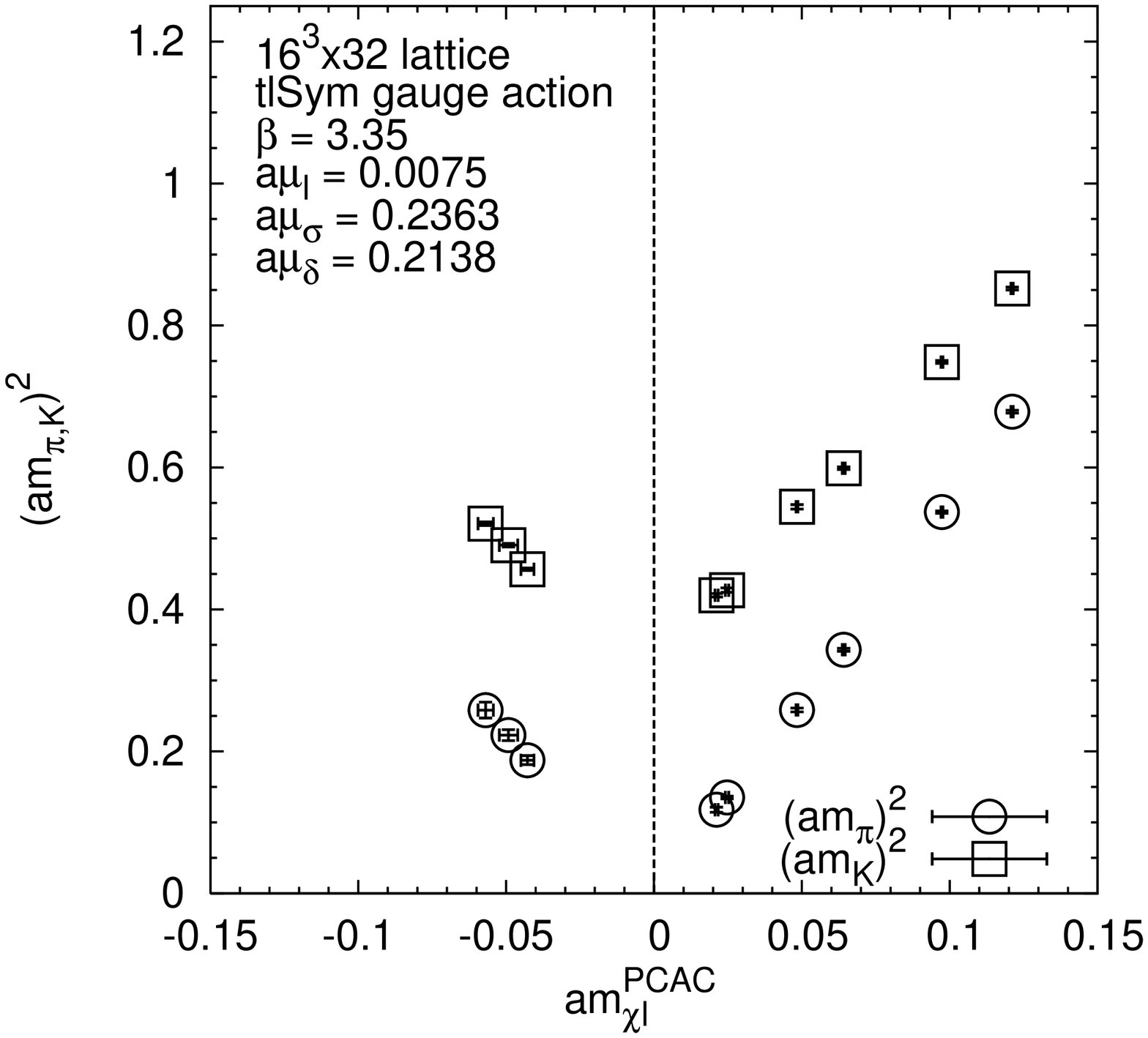}
\includegraphics[angle=-90, width=.32\textwidth]{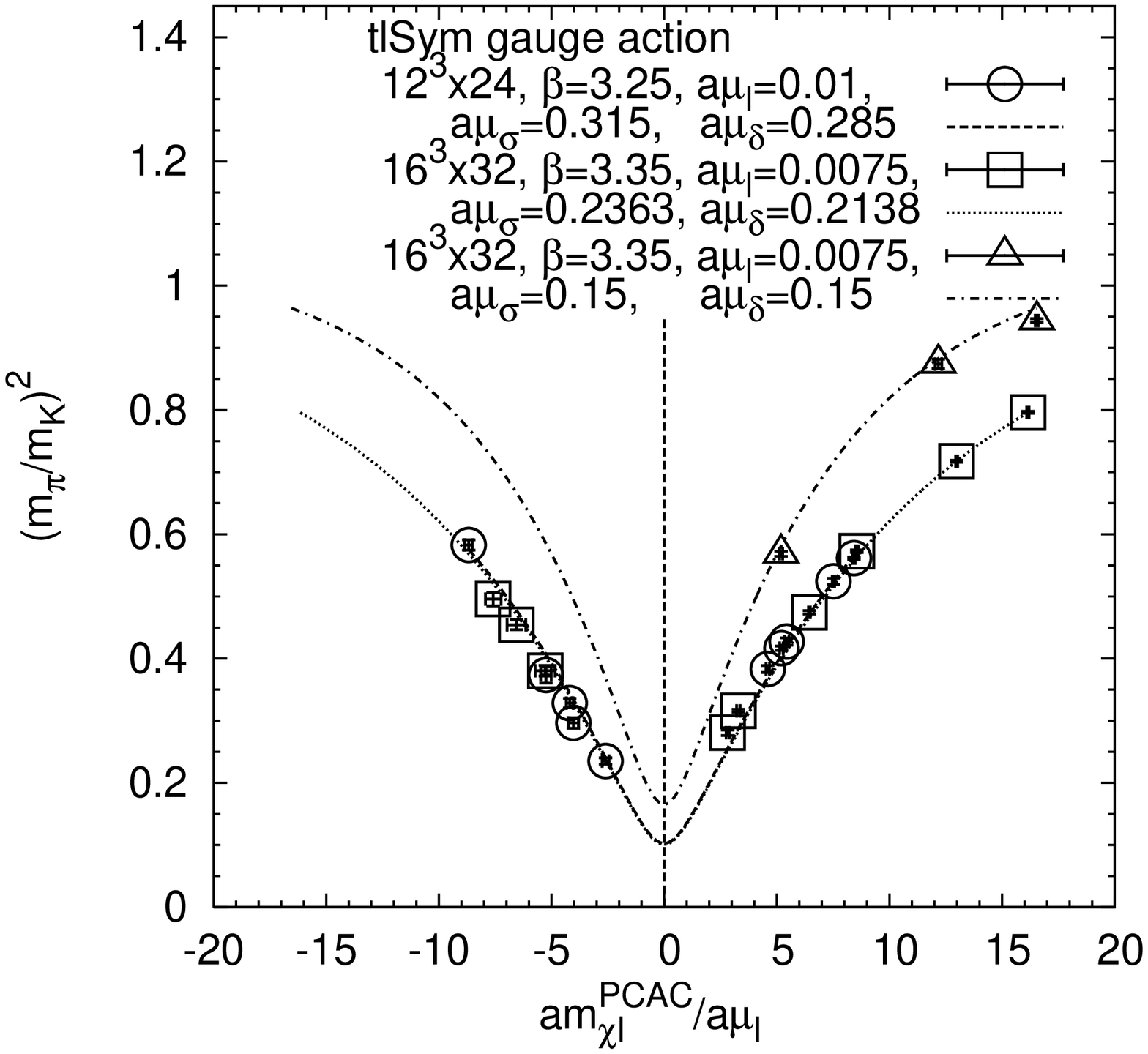}
\end{center}
\vspace*{-.75cm}
%\vspace*{-.95cm}
\caption{\label{fig:physRes}Squared pion- and kaon-masses and their ratio from $12^3\times24$ and $16^3\times32$ lattices, for details see text and \cite{Chiarappa:2006ae}}
%\vspace*{-.75cm}
\end{figure}

Here we summarize the results, details can be found in
\cite{Chiarappa:2006ae}. The most important finding concerns the phase
structure: at the coarser lattice spacing a strong metastability occurs (as
was already found for $N_f=2$ Wilson fermions, cf. 
\cite{Farchioni:2004us,Farchioni:2004fs,Farchioni:2005tu} and references
therein), leading to a minimal pion mass of approx.\ $670\textrm{MeV}$. At the
finer lattice spacing the metastability could not be observed. Actually, since
the lattice volume was fixed, we were not able to distinguish there between a
phase-transition or a cross-over scenario. Anyway, without a metastability a
lightest pion mass of $450\textrm{MeV}$ has been achieved. This implies that
on a $24^3\times48$ lattice with $a\simeq0.1\mathrm{fm}$ a pion mass of about
$300\mathrm{MeV}$ can probably be simulated. Fig.\ \ref{fig:physRes} shows the
squared pion- and kaon-masses and the ratio $(m_\pi/m_K)^2$ as functions of
the untwisted quark mass. The latter plot also contains chiral perturbation
theory guided fits to the data. Remarkably, the minimum value of these fits is
close to the physical point $(m_\pi/m_K)^2\simeq0.08$.
%
%%%%%%%%%%%%%%%%%%%%%%%%%%%%%%%%%%%%%%%%%%%%%%%%%%%%%%%%%%%%%%%%%%%%%%%%%%
%\section{Conclusions}

To conclude, dynamical $u$, $d$, $c$, and $s$ quarks can be simulated in the
Frezzotti-Rossi twisted mass formulation with a moderate tuning effort. The PHMC algorithm
is working fine for both light ($u$, $d$) and heavier ($c$, $s$) quarks.

%%%%%%%%%%%%%% figures
%
%
%\begin{figure}[h]
%\begin{center}
%\includegraphics[angle=-90, width=.32\textwidth]{Plots/plot_1.ps}%
%\includegraphics[angle=-90, width=.32\textwidth]{Plots/plot_2.ps}%
%\includegraphics[angle=-90, width=.32\textwidth]{Plots/plot_4.ps}%
%\end{center}
%\vspace*{-1.00cm}
%%\vspace*{-1.2cm}
%\caption{Example for polynomials with increasing order to approximate $x^{-1/2}$ in the interval $[5\cdot10^{-5}, 4.0]$ as used in a MSMB-update \cite{Montvay:2005tj}, shown is the relative deviation}
%\label{fig:polynoms}
%%\vspace*{-.65cm}
%\end{figure}
%
%\begin{figure}[h]
%\begin{center}
%\includegraphics[angle=-90, width=.32\textwidth]{Plots/plot_amChi_PiSq_KSq.12c24.ps}
%\includegraphics[angle=-90, width=.32\textwidth]{Plots/plot_amChi_PiSq_KSq.16c32.ps}
%\includegraphics[angle=-90, width=.32\textwidth]{Plots/chpt_fit.rev.ps}
%\end{center}
%\vspace*{-.75cm}
%%\vspace*{-.95cm}
%\caption{\label{fig:physRes}Squared pion- and kaon-masses and their ratio from $12^3\times24$ and $16^3\times32$ lattices, for details see text and \cite{Chiarappa:2006ae}}
%%\vspace*{-.75cm}
%\end{figure}
%

%%%%%%%%%%%%%%%%%%%%%%%%%%%%%%%%%%%%%%%%%%%%%%%%%%%%%%%%%%%%%%%%%%%%%%%%%
%\subsection*{Acknowledgments}

\noindent\textbf{Acknowledgments} We would like to thank the members of the ETM-Coll., especially Federico Farchioni, for their contribution. The computations were done on PC-Clusters at DESY Hamburg and Zeuthen and RWTH Aachen and at the IBM-p690 \emph{(JUMP)} installation at Forschungszentrum J\"ulich. The speaker was supported by the U.S.\ Dept.\ of Energy under contract DE-AC02-98CH10886.

%%%%%%%%%%%%%%%%%%%%%%%%%%%%%%%%%%%%%%%%%%%%%%%%%%%%%%%%%%%%%%%%%%%%%%%%%
%%%%%%%%%%%%%%%%%%%%%%%%%%%%%%%%%%%%%%%%%%%%%%%%%%%%%%%%%%%%%%%%%%%%%%%%%
%EOF

%%% Local Variables: 
%%% mode: latex
%%% TeX-master: "lattice06proc" 
%%% End:  

%%%%%%%%%%%%%%%%%%%%%%%%%%%%%%%%%%%%%%%%%%%%%%%%%%%%%%%%%%%%%%%%%%%%%%%%%%%%%%%%%%%%%%
%\begin{thebibliography}{99}
%  \bibitem{} 
%\end{thebibliography}
%%%%%%%%%%%%%%%%%%%%%%%%%%%%%%%%%%%%%%%%%%%%%%%%%%%%%%%%%%%%%%%%%%%%%%%%%%%%%%%%%%%%%

%% REFERENCES

%%% added in bib-style!!!

\bibliography{references}

\bibliographystyle{h-physrev3} %% standard is ok
%\bibliographystyle{JHEP-2} %% also puts titles, not necessary (see above)

%%%%%%%%%%%%%%%%%%%%%%%%%%%%%%%%%%%%%%%%%%%%%%%%%%%%%%%%%%%%%%%%%%%%%%%%%%%%%%%%%%%%%

\end{document}